\documentclass[conference]{IEEEtran}
\IEEEoverridecommandlockouts

\usepackage{cite}
\usepackage{amsmath,amssymb,amsfonts,amsthm}
\usepackage{mathtools}
\usepackage{algorithm}
\usepackage{algorithmic}
\usepackage{graphicx}
\usepackage{textcomp}
\usepackage{xcolor}
\usepackage{gensymb}
\usepackage{verbatim}
\def\BibTeX{{\rm B\kern-.05em{\sc i\kern-.025em b}\kern-.08em T\kern-.1667em\lower.7ex\hbox{E}\kern-.125emX}}

\newtheorem{proposition}{Proposition}

\newtheorem{lemma}{Lemma}
\newtheorem{remark}{Remark}

\renewcommand{\IEEEQED}{\IEEEQEDopen}

\begin{document}
\bstctlcite{IEEEexample:BSTcontrol}

\title{Chernoff Bounds and Saddlepoint Approximations for the Outage Probability in Intelligent Reflecting Surface Assisted Communication Systems}
\author{Tianxiong Wang,
        Gaojie Chen,~\IEEEmembership{Senior Member, IEEE,}
        Justin~P.~Coon,~\IEEEmembership{Senior Member, IEEE,}
        and Mihai-Alin~Badiu
\thanks{T. Wang, J. P. Coon and M. A. Badiu are with the Department of Engineering Science, University of Oxford, Oxford, OX1 3PJ, U.K. (e-mail: \{tianxiong.wang, justin.coon, mihai.badiu@eng.ox.ac.uk\}@eng.ox.ac.uk).}
\thanks{G. Chen is with the School of Engineering, University of Leicester, Leicester, LE1 7RH, U.K. (e-mail: gaojie.chen@leicester.ac.uk).}
}
\maketitle

\begin{abstract}
We analyze the outage probability of an intelligent reflecting surface (IRS)-assisted communication network.  A tight upper bound on the outage probability is formulated based on the Chernoff inequality. Furthermore, through an exact asymptotic (a large number of reflecting elements) analysis based on a saddlepoint approximation, we derive closed-form expressions of the outage probability for systems with and without a direct link and obtain the corresponding diversity orders. Simulation results corroborate our theoretical analysis and show the inaccuracies inherent in using the central limit theorem (CLT) to analyze system performance. Our analysis is accurate even for a small number of IRS elements in the high signal-to-noise ratio (SNR) regime.
\end{abstract}


\section{Introduction}
Recently, intelligent reflecting surfaces (IRS) have been considered as an emerging technology for the physical layer of next-generation wireless communication systems~\cite{b4}. An IRS comprises a large number of reflecting elements, which can be used to change the phases of incident waves for different functionalities, such as adding the reflected waves constructively for an intended user and destructively for other users~\cite{zr0}. Different from the traditional transmission and reception techniques,
IRS is capable of creating a controllable wireless propagation environment~\cite{di2019smart}. Compared with other related technologies, such as relaying, IRS can facilitate energy-efficient communication since, ideally, there is no power consumed by the reflecting elements~\cite{Y2019Large,Z2020AHY}. Moreover, IRS provides a full band response, which makes it a prospective technology for 5G and beyond communication systems operating in millimeter band, such as Vehicle-to-Everything (V2X) communications~\cite{Y2020R} and Internet of Things (IoT)~\cite{Q2020T}.

It has been shown that an IRS-assisted communication system can significantly improve communication performance and enhance security \cite{RS19}. Therefore, it is crucial to study the performance of IRS technology accurately. To date, many research efforts have been paid to the performance analysis of IRS-assisted networks from the perspectives of ergodic capacity, channel distribution and outage probability. For example, \cite{han2019large} and \cite{zhang2020reconfigurable} analyzed the upper bound of the ergodic spectral efficiency. The capacity degradation caused by phase adjustment errors was investigated and quantified in \cite{li2020ergodic}. As an essential metric to evaluate the reliability of the system, the outage probability has been studied in several works. For example, in \cite{jung2019reliability}, the outage probability was obtained from the distribution of the sum-rate, which was derived by using the central limit theorem (CLT) approximation. Then, the authors in \cite{b7} derived an upper bound of the outage probability of an IRS-assisted system without a direct link based on the CLT. As a step further, the work in \cite{b8} took into consideration the phase errors and showed that the channel distribution is equivalent to Nakagami fading. A more recent work \cite{Q2020P} considered a direct link between the source and destination by approximating the distribution of reflected paths as normal with CLT and computing cumulative distribution function (CDF) of the sum of a normal and a Rayleigh variables. However, the CLT is only accurate for a large number of reflecting elements and when one is interested in the distribution near the mean. As a result, the CLT can lead to significant approximation errors in the high SNR regime. To avoid these CLT issues, the authors in \cite{b9} used a gamma distribution to approximate the fading of each reflecting path. The gamma-based framework appears to offer a more accurate result. However, it provides neither a bound nor an asymptotic result for the outage probability, and the approximation accuracy cannot be quantified clearly. 

This letter provides a tight bound and asymptotic results for the outage probability of an IRS-assisted system with/without a direct link. The proposed upper bound is a Chernoff bound, whose tightness is determined by a positive parameter in the Chernoff equality. To obtain the optimal parameter, an algorithm based on the gradient descent method (GDM) was designed. On the other hand, the asymptotic expression, which is accurate for a large number of reflecting elements and high SNR, is derived based on the saddlepoint approximation. 


\section{System Model}
We focus on an IRS with \(N\) reflecting elements, labelled as \({\rm R}_n, n\in\{1, ..., N\}\), which assists the communication between a source node (S) and a destination node (D). S and D are equipped with one antenna and operate in the half-duplex mode. The IRS is assumed to work in the far-field of both S and D, therefore, the distances between S to the $R_n$ are equal for all $n$ and the distances between the $R_n$ to D are also the same for any $n$. 

Let \({h_{1n}}\), \({h_{2n}}\) and \({h_{L}}\) denote the channel coefficients for the S-to-\({\rm R}_n\), the \({\rm R}_n\)-to-D and the S-to-D channels, respectively, which are independent, circularly symmetric, complex normal random variables, each with zero mean and unit variance. It follows that the magnitudes of \({h_{1n}}\), \({h_{2n}}\) and \({h_{L}}\) follow the Rayleigh distribution with scale parameter $\sigma = \frac{1}{\sqrt{2}}$. The signal received at D can be written as
\begin{equation} \small
y = \sqrt P \left( {\sum\limits_{n = 1}^N {d_1^{-{v_1}/2}{h_{1n}}{e^{j{\theta _n}}}{d_2^{-{v_2}/2}h_{2n}} + {d_L^{-{v_L}/2}} {h_{L}}} } \right) x + u
\label{eqsys}
\end{equation} 
where $x$ is the transmitted symbol with zero mean and unit power and $P$ represents the transmit power of S. \(u \sim \mathcal{CN}(0,{\sigma_u ^2})\) denotes the additive Gaussian white noise (AWGN) received by D. $\theta _n, n\in\{1, ..., N\}$ are the phases of the reflecting elements. $d_L$, $d_1$ and $d_2$ are the distances between S to D, S to the IRS, and the IRS to D, respectively. $v_L$, $v_1$ and $v_2$ are the corresponding path loss coefficients. Each reflecting element is assumed to have unit reflection coefficient.

\subsection{Perfect Phase Alignment without the Direct Link}
If the direct link is blocked by obstacles, such as trees and buildings, which is more likely to happen when the system operates at high frequencies, the received SNR can reach its maximum when $\theta_n = - \arg ({h_{1n}}) - \arg ({h_{2n}})$, and \eqref{eqsys} can be rewritten as
\begin{equation} \small
    y_1 = H_1 \sqrt{P}d_1^{-v_1/2}d_2^{-v_2/2} x + u
\end{equation}
where 
\begin{equation} \small
    {H_1} = \sum\limits_{n = 1}^N \left| {{h_{1n}}} \right|\left| {{h_{2n}}}\right|.
\end{equation}
Denoting $\beta_R = d_1^{-v_1}d_2^{-v_2}$, the received SNR is
\begin{equation} \small
{{\gamma}_1} =\frac{{P}}{{\sigma _u^2}}{H_1 ^2}\beta_R = {H_1 ^2}{\gamma _t}\beta_R
\end{equation}
where \({\gamma _t = P/\sigma_u^2}\) is the transmit SNR.

\subsection{Perfect Phase Alignment with the Direct Link}
When the direct link is present, the phase of the $n$th reflecting path should be aligned to the phase of the direct link, i.e., \({\theta _n} = \arg ({h_{L}}) - \arg ({h_{1n}}) - \arg ({h_{2n}})\). Thus, the received signal can be expressed as
\begin{equation} \small
    y_2 = {H_2}{e^{j\arg ({h_{L}})}}{\sqrt{P\beta_R}}x + u
\end{equation}
where $H_2$ denotes the composite channel. Denoting $\alpha_L = d_L^{-v_L}/{\beta_R}$, $H_2$ can be given by
\begin{equation} \small
{H_2} = \sum\limits_{n = 1}^N {\left| {{h_{1n}}} \right|\left| {{h_{2n}}} \right| + {\sqrt \alpha_L} \left| {{h_{L}}} \right|}
\label{rs3}
\end{equation}
and the corresponding received SNR is \({\gamma _2} = {H_2^2}{\gamma _t}{\beta_R}\).

\section{Outage Probability Analysis}
The outage probability is an important measure used to evaluate the reliability of the system, which is defined as 
\begin{equation} \small
    {{{P}}^{(i)}_{\text{out}}}(\overline{\gamma}) = {\mathbb{P}}\left( {{\gamma_i} < {\overline{\gamma}}} \right) = {F_{{\gamma_i}}}\left( {{\overline{\gamma}}} \right)
\label{out1}
\end{equation}
where \(F_{\gamma_i}(\cdot)\) stands for the CDF of $\gamma_i$; $\overline{\gamma}$ denotes the threshold SNR; $i\in\{1,2\}$ corresponds to the scenarios without and with the direct link, respectively. The outage probability for a certain threshold $\overline{\gamma}$ is equivalent to the probability that the channel coefficient falls below a value related to $\overline{\gamma}$, i.e.,
\begin{equation} \small
    {{{P}}^{(i)}_{\text{out}}}(\overline{\gamma}) = F_{H_{i}}\left(\sqrt{\frac{\overline{\gamma}}{\gamma_t \beta_R}} \right).
    \label{ou1}
\end{equation}
Based on the outage probability, we can write the diversity order of the system as
\begin{equation} \small
    {d}_{i} = \mathop {\lim }\limits_{{\gamma _t} \to \infty }  - \frac{{\log  {{P^{(i)}_{\text{out}}}} }}{{\log {\gamma _t}}}
\label{d2}.
\end{equation}
Therefore, in what follows, we will focus on the CDF of the channel coefficients.

\subsection{Perfect Phase Alignment without the Direct Link}
\subsubsection{Chernoff Bound}
Let $G_n = \left| {{h_{1n}}} \right|\left| {{h_{2n}}}\right|$. Since $\left| {{h_{1n}}} \right|$ and $\left| {{h_{2n}}} \right|$ follow the independent Rayleigh distribution with the parameter $\sigma = \frac{1}{\sqrt{2}}$, the probability density function (PDF) of $G_n$ is \cite{b10}
\begin{equation} \small
    f_{{G_n}}(x) = 4x{K_0}(2x).
\end{equation}
A Chernoff upper bound on the CDF of $H_1$, denoted as $F_{H_1}(s)$, can be written as \cite{chernoff1952measure}
\begin{equation} \small
    F_{H_{1}}\left(s\right) \leq \underset{t>0}{\min}\ e^{ts} \prod_{n=1}^{N}{\rm E}\left[ e^{-tG_n} \right]
\end{equation}
where ${\rm E}\left[ e^{-tG_n} \right]$ can be calculated to be 
\begin{equation} \small
    {\rm E}\left[ e^{-tG_n} \right] = \left\{
    \begin{split}
        \frac{1}{1-\frac{t^2}{4}} &- \frac{t \arccos\left(\frac{t}{2}\right)}{2 \left(1-\frac{t^2}{4}\right)^{3/2}}, &t>0\,\&\,t\neq 2\\
        &\frac{1}{3}, &t=2.~~~~~~~~~
    \end{split}
    \right.
    \label{ex1}
\end{equation}
Denoting $w(t) = e^{ts} \prod_{n=1}^{N}{\rm E}\left[ e^{-tG_n} \right]$, we are now in the position to find the optimal value of $t$ which can minimize $w\left(t\right)$. Fortunately, $w\left(t\right)$ is proved to be convex in $t$. 
\begin{lemma}
    $w(t)$ is convex in $t$.
\end{lemma}
\begin{IEEEproof}
    Since $G_n$, $n=1,2,...,N$ are independent random variables, $w(t)$ can be rewritten as
    \begin{equation} \small
        \begin{split}
            w(t) &= e^{ts} {\rm E}\left[e^{-t\sum\limits_{n = 1}^{N}G_n}\right]\\
            & = \int_0^{+\infty } f_{H_1}(x) e^{t \left(s-x \right)} \, {\rm d}x
        \end{split}
        \label{pr}
    \end{equation}
where $f_{H_1}(x)$ denotes the probability density function (PDF) of $H_1$. Due to the fact that $e^{t \left(s-x \right)}$ is convex in $t$ and $f_{H_1}(x) \geq 0$, $w(t)$ is convex in $t$ \cite[sec.~3.2]{boyd2004convex}.
\end{IEEEproof}
\begin{remark}
    It should be noted that the convexity of $w(t)$ does not depend on the specific distribution of $G_n$. In other words, we can arrive at the conclusion of convexity as long as $G_n$, $n=1,2,...,N$ are independent.
\end{remark}
It is mathematically intractable to obtain a close-form expression of the minimum of $w(t)$. Alternatively, we propose the following gradient descent algorithm to find an approximated numerical result.
\begin{algorithm}
\caption{Gradient Descent Method for Minimizing $w\left(t\right)$}
\label{alg1}
\begin{algorithmic}[1]
\REQUIRE $s$, $N$;
\ENSURE Minimum of $w(t)$;
\IF {$w'\left(0\right) \geq 0$}
\STATE $\underset{t>0}{\min}\ w\left(t\right) = 1$;
\ELSE
\STATE Initialize a starting point $t_{0}>0$;
\STATE In the $i$th iteration, determine a descent direction \[\Delta t_{i} = - w'(t_{i-1}),\] and update $t_{i}$ by:
\begin{equation}\nonumber \small
    t_i = t_{i-1} + \xi_{i} \Delta t_{i},
\end{equation}
where $\xi_{i}$ denotes the step size, which is calculated by the backtracking line search method;
\STATE Go back to 5 until the stopping criterion $\left|w'(t_{i})\right| \leq \varepsilon$ is satisfied, where $\varepsilon$ is the tolerance;
\STATE $\underset{t>0}{\min}\ w\left(t\right) = w\left(t_{i}\right)$;
\ENDIF
\end{algorithmic}
\end{algorithm}

\subsubsection{Saddlepoint Approximation}
We again treat the case of perfect phase adjustment without a direct link. As defined before, $G_n = |h_{1n}||h_{2n}|$. We require the distribution of ${H_1} = \sum\limits_{n = 1}^N G_n$ for large $N$.  Moreover, our interest lies in the tails of the distribution where the central limit approximation is not accurate.  Hence, we resort to a saddlepoint approximation.  The result is summarised in the following proposition.

\begin{proposition}
  The cumulative distribution function of $H_1$ obeys the asymptotic equivalence
  \begin{equation} \small
      \begin{split}
           F_{H_1}(s) \sim &\frac{2^{N}N!}{N^{N} \sqrt{4\pi N}}e^{-2N\left(\ln \left(\frac{2N}{s} \right)-1\right)} \\
      &\times\sum_{n=0}^N\frac{\left(2N\left(\ln \left(\frac{2N}{s}\right)-1\right)\right)^{n}}{n!},\quad N\to\infty.
      \end{split}
  \end{equation}
  
\end{proposition}
\begin{IEEEproof}
    We give a proof based on the saddlepoint approximation.  
    Let $L_{H_{1}}(t) = {\rm E}[e^{-t H_{1}}]$ represent the Laplace transform of the PDF of $H_1$, denoted as $f_{H_{1}}(x)$.  This is well defined due to the fact that $G_n \geq 0$ for $n = 1,\ldots,N$ and that fact that the Laplace transform of the PDF corresponding to the random variable $G_n$, denoted by $L_G(t)$, exists.  Indeed, we can write $L_{H_{1}}(t) = L_G(t)^N$ due to independence.  
    
    Now, the probability density function of $H_1$ can be written as
    \begin{equation} \small
        f_{H_{1}}(x) = \frac{1}{2\pi i} \int_{c-i\infty}^{c+i\infty} e^{N \lambda(t)} {\rm d}t
    \end{equation}
    where $c$ is greater than the real part of the singularities of the integrand and
    \begin{equation} \small
        \lambda(t) = \ln L_G(t) + \frac{x}{N}t.\label{lambdat}
    \end{equation}
    The integral is dominated for large $N$ by the saddlepoint of $\lambda(t)$.  Denote this saddlepoint of $\lambda(t)$ by $\hat t$.  We have that
    \begin{equation} \small
        Q(\hat t) \coloneqq -\frac{L_G'(\hat t)}{L_G(\hat t)} = \frac x N
    \end{equation}
    must hold.  For large $N$, we must have that $\hat t$ is large, since $L_G'(\hat t) \approx 0$.  Hence, we expand $Q(t)$ for a large argument with positive real part.  It is straightforward to show that
    \begin{equation} \small
        Q(t) \sim \frac 2 t,\qquad \Re\{t\}\to\infty.
    \end{equation}
    It follows that
    \begin{equation} \small
        \hat t \sim \frac{2 N}{x},\qquad N\to\infty.
        \label{hatt}
    \end{equation}
    By considering the second derivative of $\lambda(t)$ evaluated at large $t$, it is possible to show that, indeed, a maximum occurs at $\hat t$.
    
    Deforming the integral contour to pass $\hat t$ and using the approximation $\lambda(t) \approx \lambda(\hat t) + (t - \hat t)^2 \lambda''(\hat t)/2$, we have that
    \begin{equation} \small
        f_{H_{1}}(x) \approx \frac{1}{2\pi i} \int_{c-i\infty}^{c+i\infty} e^{N (\lambda(\hat t) + (t - \hat t)^2 \lambda''(\hat t)/2)} {\rm d}t.
    \end{equation}
    Note that this approximation is asymptotically precise (in $N$).  Rearranging, substituting variables ($t - \hat t = re^{i\phi}$), and evaluating the Gaussian integral yields
    \begin{equation} \small
        f_{H_{1}}(x) \sim \frac{e^{N\lambda(\hat t)}}{\sqrt{2\pi N \left|\lambda''(\hat t)\right|}}
    \end{equation}
    in the usual way for saddlepoint approximations.  One can evaluate
    \begin{equation} \small
        \lambda(\hat t) \sim 2 - 2 \ln\frac{N}{x}+\ln\left(\ln\left(\frac{2 N}{x}\right) - 1\right)
    \end{equation}
    and
    \begin{equation} \small
        \lambda''(\hat t) \sim \frac{x^2}{2N^2}.
    \end{equation}
    Hence,
    \begin{equation} \small
        f_{H_{1}}(x) \sim \frac{e^{2N}\left(\ln\left( \frac{2N}{x}\right)-1\right)^N}{\left(\frac{N}{x}\right)^{2N-1}\sqrt{\pi N}},\quad N\to\infty.
        \label{fh124}
    \end{equation}
    By integrating \eqref{fh124} directly, we arrive at
    \begin{equation} \small
        F_{H_1}\left(s\right) \sim \frac{2^{N}}{(N)^{N}\sqrt{4\pi N}}\Gamma\left(N+1,2N\left(\ln \left(\frac{2N}{s} \right)-1\right)\right).
    \end{equation}
    The stated asymptotic equivalence for the cumulative distribution then follows from the properties of the upper incomplete gamma function.
\end{IEEEproof}

\begin{remark}
Based on \eqref{ou1} and \eqref{d2}, the diversity order of the system without a direct link can be calculated by
\begin{equation} \small
    d_1 = N.
\end{equation}
\end{remark}

\begin{remark}
It is worth highlighting that this asymptotic expression is a refined large deviation result.  In fact, to leading order, we have that
\begin{equation} \small
    F_{H_1}\left(s\right) \sim \frac{4^{N}e^{-2N\left(\ln \left(\frac{2N}{s} \right)-1\right)}}{\sqrt{4\pi N}}{\left(\ln \left(\frac{2N}{s} \right)-1\right)}^{N}
    \label{FH1a}
\end{equation}
from which we observe that the rate function written for large, finite $N$ is
\begin{equation} \small
    \begin{split}
        J_N(s) &\sim -\frac 1 N \ln F_{H_1}\left(s\right) \\
        &\sim -\ln4 + 2\left(\ln\left(\frac{2N}{s}\right)-1\right)\\
        &~~~- \ln\left(\ln \left(\frac{2N}{s}\right) - 1\right) + \frac{\ln4\pi N}{2N}.
    \end{split}
\end{equation}
Applying the large deviation principle directly (instead of the saddlepoint approximation) yields all but the $O\left(\frac{\ln N}{N}\right)$ term for the finite rate function. By using the explicit saddlepoint approximation, however, we obtain additional information about the exponential decay of the distribution as $N$ grows large.  Furthermore, we see that, crucially, the asymptotic distribution is accurate for \textit{finite} $N$ as long as $N \gg s$.  In practice, we will be interested in values of $s$ that are inversely proportional to the average received SNR at the destination node.  Hence, this condition will be met for large enough SNR.  Conversely, when the SNR is on the order of $1/N$, we can resort to a central limit approximation, or perhaps a refined Edgeworth expansion, for the channel distribution.
\end{remark}

\subsection{Perfect Phase Alignment with the Direct Link}

\subsubsection{Chernoff Bound}
We now turn to the case of perfect phase alignment with a direct link. Let $D = {\sqrt \alpha_L} \left| {{h_{L}}} \right|$. According to the system model, D follows the Rayleigh distribution with the scale parameter $\sigma_D = \sqrt{\frac{\alpha_L}{2}}$. A Chernoff upper bound of $F_{H_2}\left(s\right)$ can be written as
\begin{equation} \small
    F_{H_2}\left(s\right) \leq \underset{t>0}{\min}\ e^{ts} {\rm E}\left[ e^{-tD} \right] \prod_{n=1}^{N}{\rm E}\left[ e^{-tG_n} \right]
\end{equation}
where ${\rm E}\left[ e^{-tD} \right]$ can be calculated to be
\begin{equation} \small
    {\rm E}\left[ e^{-tD} \right] = 1 - \frac{\sqrt{\pi \alpha_L}}{2} t e^{\frac{\alpha_L t^2}{4}} \text{erfc}\left(\frac{\sqrt{\alpha_L} t}{2}\right).
\end{equation}
Let $z(t) = e^{ts} {\rm E}\left[ e^{-tD} \right] \prod_{n=1}^{N}{\rm E}\left[ e^{-tG_n} \right]$. Since $z(t)$ can be rewritten as
\begin{equation} \small
    z(t) = e^{ts} {\rm E}\left[e^{-t\left(D + \sum\limits_{n = 1}^{N}G_n\right)} \right]
\end{equation}
we can prove the convexity of $z(t)$ following the same procedure in Proposition 1. Moreover, the optimal value of $t$ for minimizing $z(t)$ can be obtained efficiently by the gradient descent method. 

\subsubsection{Saddlepoint Approximation}
The Laplace transform of $f_D(s)$ is
    \begin{equation} \small
        L_D\left(t\right) = 1 - \frac{\sqrt{\pi \alpha_L}}{2} t e^{\frac{\alpha_L t^2}{4}} \text{erfc}\left(\frac{\sqrt{\alpha_L} t}{2}\right).
    \end{equation}
Following this, we can write the PDF of $H_2$ as
    \begin{equation} \small
        f_{H_2}(x) = \frac{1}{2\pi i} \int_{c-i\infty}^{c+i\infty} L_D\left(t\right)e^{N \lambda(t)} {\rm d}t 
        \label{fh2}
    \end{equation}
where $c$ is greater than the real part of the singularities of the integrand and $\lambda(t)$ is defined in \eqref{lambdat}. The integral in \eqref{fh2} is dominated for large $N$ by $\hat t$, which is the saddlepoint of $\lambda(t)$ given in \eqref{hatt}. Invoking the saddlepoint approximation by setting
\begin{equation} \small
        L_{D}(t) \approx L_{D}(\hat t) \quad {\rm and} \quad \lambda(t) \approx \lambda(\hat t) + (t - \hat t)^2 \frac{\lambda''(\hat t)}{2}
\end{equation}
and evaluating the integral \eqref{fh2}, we can arrive at 
    \begin{equation} \small
        \begin{split}
            f_{H_2}(x) &\sim L_D\left(\hat t\right) \frac{e^{n\lambda(\hat t)}}{\sqrt{2\pi n \lambda''(\hat t)}} \\
            &\sim \left(1-\frac{\sqrt{\pi \alpha_L} N  e^{\frac{\alpha_L N^2}{x^2}} \text{erfc}\left(\frac{\sqrt{\alpha_L} N}{x}\right)}{x}\right)\\
            &~~~~\times \frac{e^{2N}\left(\ln\left( \frac{2N}{x}\right)-1\right)^N}{\left(\frac{N}{x}\right)^{2N-1}\sqrt{\pi N}},\quad N\to\infty.
        \end{split}
        \label{fH2a}
    \end{equation}
Based on the asymptotic expression of the complementary error function for large argument \cite{olver1997asymptotics}, we have
\begin{equation} \small
    \text{erfc}\left(\frac{\sqrt{\alpha_L} N}{x}\right) \sim \frac{x e^{-\frac{\alpha_L N^2}{x^2}} \left(1-\frac{x^2}{2 \alpha_L N^2}\right)}{\sqrt{\pi \alpha_L} N},\quad N\to\infty.
    \label{erfc1}
\end{equation}
Taking \eqref{erfc1} into \eqref{fH2a} and integrating \eqref{fH2a} directly, we obtain the CDF of $H_2$:
\begin{equation} \small
        F_{H_2}\left(s\right) \sim \frac{2^N \sqrt{N} \Gamma \left(N+1,2 (N+1) \left(\ln \left(\frac{2 N}{s}\right)-1\right)\right)}{e^2 \sqrt{\pi } {\alpha_L} (N+1)^{N+1}},\quad N\to\infty.
       \label{eq37}
\end{equation}
According to \eqref{ou1} and \eqref{d2}, the diversity order can then be derived as
\begin{equation} \small
    d_2 = N + 1.
\end{equation}

\section{Numerical Results}
In this section, we investigate the outage probability of the IRS-assisted communication systems and evaluate our proposed upper bounds and asymptotic results by Monte Carlo (MC) simulations. We set $\overline{\gamma} = 0\,{\rm dB}$, $d_1 = 5\,{\rm m}$, $d_2 = 5\,{\rm m}$, $d_L = 7\,{\rm m}$, $v_1 = 2.5$, $v_2 = 2.5$, $v_L = 3.5$. 

\begin{figure}[!]
    \centerline{\includegraphics[scale=0.46]{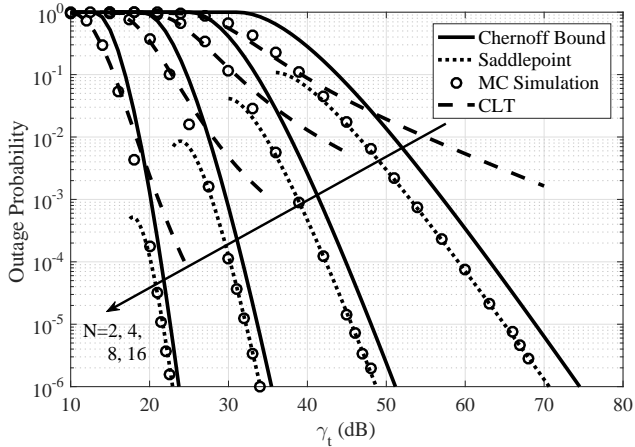}}   
    \caption{Outage probability versus transmit SNR without the direct link for different $N$.}
    \label{figcher1}
\end{figure}

Fig. \ref{figcher1} illustrates the outage probability versus the transmit SNR $\gamma_t$ when the direct link is blocked. The Chernoff upper bounds and the asymptotic results based on the saddlepoint approximation are compared with MC simulations and the CLT approximation. It can be seen that the Chernoff bounds are more accurate than the CLT method in the high transmit SNR regime. For example, when $N=16$, the Chernoff bound is more accurate for $\gamma_t > 20\,{\rm dB}$. We can also observe that increasing the number of reflecting elements can significantly reduce the transmit SNR to achieve a given outage probability as expected. For example, when the outage probability is $10^{-4}$, the transmit SNR is $30\,{\rm dB}$ for $N=8$ and $20\,{\rm dB}$ for $N=16$. Since the asymptotic behavior of \eqref{FH1a} will appear at very low outage probability for large $N$, the largest $N$ we use in analysis is $N=16$. It is shown that the CLT approximation and the saddlepoint approximation match nicely with the simulation results in the low and high transmit SNR regimes, respectively. This confirms that the saddlepoint approximation is valid in the tails of the distribution where the CLT method is not accurate.

\begin{figure}[!]
    \centerline{\includegraphics[scale=0.46]{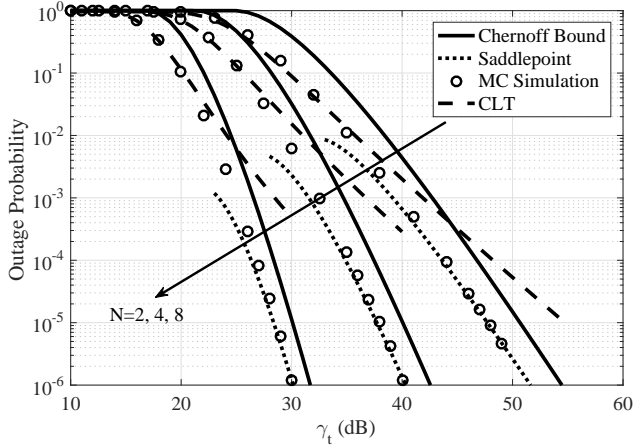}}   
    \caption{Outage probability versus transmit SNR with the direct link for different $N$.}
    \label{figs1}
\end{figure}

Fig. \ref{figs1} shows the outage probability of the system with a direct link. Similar to the case without a direct link, it can be seen that the Chernoff bounds and the asymptotic results based on the saddlepoint approximation are close to the MC results.
Comparing Figs. \ref{figcher1} and \ref{figs1}, we can observe a significant performance gain due to the existence of a direct link. 

\section{Conclusions}
In this letter, the outage probability of IRS-assisted communication systems was investigated. It was shown that the Chernoff upper bound is tighter than the CLT approximation for high transmit SNR. Besides, although the asymptotic analysis is for large  $N$, numerical results confirmed that the reported asymptotic expressions are still accurate for small $N$ in the high transmit SNR regime. We also demonstrated that increasing the number of reflecting elements can significantly reduce the outage probability for the same transmit SNR.



\bibliographystyle{IEEEtran}
\bibliography{irs}

\begin{thebibliography}{10}
\providecommand{\url}[1]{#1}
\csname url@samestyle\endcsname
\providecommand{\newblock}{\relax}
\providecommand{\bibinfo}[2]{#2}
\providecommand{\BIBentrySTDinterwordspacing}{\spaceskip=0pt\relax}
\providecommand{\BIBentryALTinterwordstretchfactor}{4}
\providecommand{\BIBentryALTinterwordspacing}{\spaceskip=\fontdimen2\font plus
\BIBentryALTinterwordstretchfactor\fontdimen3\font minus
  \fontdimen4\font\relax}
\providecommand{\BIBforeignlanguage}[2]{{%
\expandafter\ifx\csname l@#1\endcsname\relax
\typeout{** WARNING: IEEEtran.bst: No hyphenation pattern has been}%
\typeout{** loaded for the language `#1'. Using the pattern for}%
\typeout{** the default language instead.}%
\else
\language=\csname l@#1\endcsname
\fi
#2}}
\providecommand{\BIBdecl}{\relax}
\BIBdecl

\bibitem{b4}
C.~Liaskos, S.~Nie \emph{et~al.}, ``A new wireless communication paradigm
  through software-controlled metasurfaces,'' \emph{IEEE Commun. Mag.},
  vol.~56, no.~9, pp. 162--169, 2018.

\bibitem{zr0}
Q.~Wu and R.~Zhang, ``Intelligent reflecting surface enhanced wireless network:
  Joint active and passive beamforming design,'' in \emph{Proc. IEEE GLOBECOM},
  Abu Dhabi, United Arab Emirates, Dec. 2018, pp. 1--6.

\bibitem{di2019smart}
M.~Di~Renzo, M.~Debbah \emph{et~al.}, ``Smart radio environments empowered by
  reconfigurable ai meta-surfaces: An idea whose time has come,'' \emph{EURASIP
  J. Wireless Commun. Net.}, vol. 2019, no.~1, pp. 1--20, 2019.

\bibitem{Y2019Large}
Y.~{Liang}, R.~{Long} \emph{et~al.}, ``Large intelligent surface/antennas
  ({LISA}): Making reflective radios smart,'' \emph{J. Commun. Inf. Net.},
  vol.~4, no.~2, pp. 40--50, 2019.

\bibitem{Z2020AHY}
Z.~{Abdullah}, G.~{Chen} \emph{et~al.}, ``A hybrid relay and intelligent
  reflecting surface network and its ergodic performance analysis,'' \emph{IEEE
  Wireless Commun. Lett.}, to appear.

\bibitem{Y2020R}
Y.~{Chen}, Y.~{Wang} \emph{et~al.}, ``Resource allocation for intelligent
  reflecting surface aided vehicular communications,'' \emph{IEEE Trans. Veh.
  Technol}, pp. 1--1, 2020.

\bibitem{Q2020T}
Q.~{Wu} and R.~{Zhang}, ``Towards smart and reconfigurable environment:
  Intelligent reflecting surface aided wireless network,'' \emph{IEEE Commun.
  Mag.}, vol.~58, no.~1, pp. 106--112, 2020.

\bibitem{RS19}
X.~Yu, D.~Xu \emph{et~al.}, ``Robust and secure wireless communications via
  intelligent reflecting surfaces,'' \emph{IEEE J. Select. Areas Commun.}, to
  appear.

\bibitem{han2019large}
Y.~Han, W.~Tang \emph{et~al.}, ``Large intelligent surface-assisted wireless
  communication exploiting statistical {CSI},'' \emph{IEEE Trans. Veh.
  Technol}, vol.~68, no.~8, pp. 8238--8242, 2019.

\bibitem{zhang2020reconfigurable}
H.~Zhang, B.~Di \emph{et~al.}, ``Reconfigurable intelligent surfaces assisted
  communications with limited phase shifts: How many phase shifts are enough?''
  \emph{IEEE Trans. Veh. Technol}, vol.~69, no.~4, pp. 4498--4502, 2020.

\bibitem{li2020ergodic}
D.~Li, ``Ergodic capacity of intelligent reflecting surface-assisted
  communication systems with phase errors,'' \emph{IEEE Commun. Lett.}, 2020.

\bibitem{jung2019reliability}
M.~Jung, W.~Saad \emph{et~al.}, ``Reliability analysis of large intelligent
  surfaces ({LIS}s): Rate distribution and outage probability,'' \emph{IEEE
  Wireless Commun. Lett.}, vol.~8, no.~6, pp. 1662--1666, 2019.

\bibitem{b7}
D.~{Kudathanthirige}, D.~{Gunasinghe}, and G.~{Amarasuriya}, ``Performance
  analysis of intelligent reflective surfaces for wireless communication,'' in
  \emph{Proc. IEEE ICC}, Dublin, Ireland, Jun. 2020, pp. 1--6.

\bibitem{b8}
M.~{Badiu} and J.~P. {Coon}, ``Communication through a large reflecting surface
  with phase errors,'' \emph{IEEE Wireless Commun. Lett.}, vol.~9, no.~2, pp.
  184--188, 2020.

\bibitem{Q2020P}
Q.~{Tao}, J.~{Wang}, and C.~{Zhong}, ``Performance analysis of intelligent
  reflecting surface aided communication systems,'' \emph{IEEE Commun. Lett.},
  pp. 1--1, 2020.

\bibitem{b9}
S.~{Atapattu}, R.~{Fan} \emph{et~al.}, ``Reconfigurable intelligent surface
  assisted two–way communications: Performance analysis and optimization,''
  \emph{IEEE Trans. Wireless Commun.}, pp. 1--1, 2020.

\bibitem{b10}
J.~Salo, H.~M. El-Sallabi, and P.~Vainikainen, ``The distribution of the
  product of independent rayleigh random variables,'' \emph{IEEE Trans.
  Antennas Propag.}, vol.~54, no.~2, pp. 639--643, 2006.

\bibitem{chernoff1952measure}
H.~Chernoff \emph{et~al.}, ``A measure of asymptotic efficiency for tests of a
  hypothesis based on the sum of observations,'' \emph{Ann. Math. Statist.},
  vol.~23, no.~4, pp. 493--507, 1952.

\bibitem{boyd2004convex}
S.~Boyd and L.~Vandenberghe, \emph{Convex Optimization}.\hskip 1em plus 0.5em
  minus 0.4em\relax Cambridge university press, 2004.

\bibitem{olver1997asymptotics}
F.~Olver, \emph{Asymptotics and special functions}.\hskip 1em plus 0.5em minus
  0.4em\relax CRC Press, 1997.

\end{thebibliography}

\end{document}